# Dielectric Anomaly at $T_N$ in LaMnO$_3$

Parthasarathi Mondal,[1] Dipten Bhattacharya,[1*] Pranab Choudhury,[1] and Prabhat Mandal[2]

[1]*Sensor & Actuator Division, Central Glass and Ceramic Research Institute, Kolkata 700 032, India*
[2]*Saha Institute of Nuclear Physics, 1/AF Bidhannagar, Kolkata 700 064, India*

We observe a distinct anomaly in dielectric permittivity ($\varepsilon'$) as well as relaxation time ($\tau_0$) versus temperature ($T$) pattern at the antiferromagnetic transition point $T_N$ in single crystal of LaMnO$_3$. The equivalent circuit analysis of the impedance spectra across $T_N$ reveals clear anomaly in the capacitive component $C_0$ at $T_N$. Since no structural transition takes place across $T_N$, the anomaly in $\tau_0$ and $C_0$ at $T_N$ possibly signifies multiferroicity stemming from coupling between orbital and spin order in LaMnO$_3$.

The observation of multiferroicity with strong coupling between electric polarization (**P**) and magnetization (**M**) in perovskite manganites has triggered an intense search for different systems exhibiting such coupling.[1] The presently known multiferroics could be broadly classified into the ones with purely electronic multiferroicity and the ones where mutiferroicity results from displacive ferroelectricity driven either directly by magnetic frustration or by covalency in a sublattice different from the magnetic sublattice. Obviously, when ferroelectricity and magnetization originate in different sublattices, one observes very weak coupling between **P** and **M**. On the other hand, when the onset of incommensurate spiral magnetic order at $T_{IC}$ ($<T_N$) in perovskite TbMnO$_3$ breaks simultaneously both spatial and time inversion symmetries, it exhibits a strong coupling between **P** and **M**. Likewise, the electronic multiferroicity too, being explored in strongly correlated electron systems,[2,3] might exhibit strong coupling as one can maneuver it via fine interplay among charge, spin, and orbital degrees of freedom. Therefore, it is worth searching for different strongly correlated electron systems for electronic multiferroicity where the ferroelectricity is driven by strong correlation as opposed to covalency. It is both interesting and surprising, in this context, that we observe a clear dielectric anomaly at the magnetic transition point ($T_N$) even in pure LaMnO$_3$. This is an essential signature of multiferroicity. The origin of the reasonably high electric polarization – static dielectric constant $\varepsilon_0$ ~18-20 [Ref. 4] – in pure LaMnO$_3$ is not quite clear at this moment. It could either originate from orbital order driven asymmetry in Mn3$d$-O2$p$-Mn3$d$ electron overlap resulting from strong correlation (purely electronic effect) or from semi-covalency (coupled electron-lattice effect), also resulting from orbital order. The presence of a role of orbital order in the observed electric polarization offers a natural explanation to the dielectric anomaly at $T_N$ via spin-orbital order coupling. The theory shows[5,6] that A-type antiferromagnetic order in LaMnO$_3$, observed below $T_N$ (~140 K), is driven by C-type orbital order while the first direct experimental evidence of interplay between spin and orbital degrees of freedom is gathered in pseudocubic perovskite KCuF$_3$ [Ref. 7] using the intensity of orbital Bragg peaks in resonant X-ray scattering (RXS) experiment.

The low frequency (10 Hz – 10 MHz) dielectric properties have been measured on high quality single crystal of LaMnO$_3$ across a temperature regime 77-200 K. The single crystal has been grown from solid ingots of LaMnO$_3$ by floating zone technique in an image furnace under flowing Ar. The crystals were characterized by powder X-ray diffraction and Laue photography. The bulk ingots have been prepared by sintering powder of LaMnO$_3$ in inert atmosphere. The details of the sample preparation and their characterization are available in our earlier papers.[8-11] The dielectric properties across a frequency range 10 Hz – 10 MHz have been measured in Solartron Dielectric Interface (Model 1296) coupled with a Frequency Response Analyzer (Model 1260). The cryostat of Oxford instrument was used for the low temperature measurements.

In Fig. 1, we show the real part of the permittivity [$\varepsilon'(\omega,T)$] versus temperature ($T$) patterns across $T_N$ as well as across a wide temperature range 77-900 K in the inset (a). The anomalies at $T_N$ (~150 K) and $T_{JT}$ (~750 K) are conspicuous. In the inset (b), we also show a representative polarization ($P$) versus electric field ($E$) loop along with the time ($t$) dependence of polarization. The impedance spectra ($Z'$, $Z''$) at different temperatures across $T_N$ are shown in Fig. 2a. The relaxation feature is evident in the spectra, as the peak in $Z''$ versus frequency plot shifts toward higher frequency with temperature. In Fig. 2b and 2c, we show the impedance spectra on complex plane and the corresponding equivalent circuit. Finally, in Fig. 3 we show the circuit elements $R_N$ and $C_0$, corresponding to the impedance spectra, as a function of temperature. Distinct anomaly could be seen in $C_0$ at $T_N$. The relaxation time scale ($\tau_0$) versus temperature plot too, shown in Fig. 3d, depicts clear anomaly at $T_N$.

The entire $\varepsilon'$-$T$ pattern across 77-900 K [Fig. 1 inset (a)] has two prominent features: (i) near $T_N$ and (ii) near $T_{JT}$. Far below $T_N$, $\varepsilon'$ is nearly temperature- and frequency-independent, as expected. Following the anomaly at $T_N$, $\varepsilon'$ rises with $T$. Finally, $\varepsilon'$ becomes nearly temperature-independent beyond $T_{JT}$. The P-E loop [Fig. 1 inset (b)] does not signify any ferroelectric order yet the time-dependence plot resembles the 'domain-switching-like' pattern.[12] Finite loop area signifies presence of irreversible local domain fluctuations. The $\varepsilon'$-$T$ pattern near $T_{JT}$ too, follows closely the defect-lattice or localized polar cell model – freezing of local dipoles below $T_{JT}$ and divergence of relaxation time – proposed in the context of electronic/structural phase transition in $CaCu_3Ti_4O_{12}$ or $La_{2-x}Sr_xCuO_4$.[13] In the mean-field approximation of the localized defect (polar) cell model, the complex $\varepsilon$ is given by $\varepsilon(\omega,T) = \dfrac{\varepsilon_0}{1-[\alpha\gamma/(\gamma-i\omega)]}$ where $\alpha$ depends on defect cell concentration $c$ and polarizability $p$ of a cell, $\alpha = (4\pi/3)cp\varepsilon_0$ and $\gamma$ is the relaxation rate of the defect cells, $\gamma = \gamma_0.\exp(-\Delta/T)$, $\Delta$ is the energy barrier of relaxation in temperature unit. The experimental $\varepsilon'$-$T$ near $T_{JT}$ can be fitted by this model [solid line in Fig. 1 inset (a)]. From these results, it appears that the electrical polarization results from local polar cells (domains) with no global ferroelectric order. Presence of fluctuating local domains tends to broaden the cross-over feature at $T_N$ and also gives rise to frequency dependence of $\varepsilon'$ above ~77 K.

We have analyzed the impedance spectra using the complex plane plot. The complex plane impedance spectra have been fitted by the generalized Davidson-Cole type relaxation equation[14]

$$Z^* = R_\infty + \frac{R_0 - R_\infty}{(1 + i\omega\tau_0)^\beta} \qquad (1)$$

where $R_0$ and $R_\infty$ are static and high frequency resistance, respectively, $\omega$ is the frequency and $\beta$ is the Kohlrausch exponent[15] which measures the width of the relaxation time scale distribution; $\beta = 1$ for purely Debye relaxation while it varies within 0 and 1 for non-Debye correlated relaxation. The Debye model is applicable to the relaxation dynamics of independent or uncorrelated dipole moments in ideal systems with very high degree of purity where as the Davidson-Cole model describes more involved relaxation process of correlated dipoles. In most of the real dielectric solids/liquids, it has been found that the Davidson-Cole model with wide distribution of relaxation time is applicable. This is because of intrinsic inhomogeneity in a real system which gives rise to local domain formation and hence broader relaxation patterns. In that case, the complex-plane plot deviates from perfect semicircle with center on the real-axis to an arc with depression of the center of the arc below the real-axis. The Kohlrausch-Williams-Watts (KWW) relaxation function assumes stretched exponential form with Kohlrausch parameter β varying within 0 to 1.0. Such non-Debye Davidson-Cole model of dielectric relaxation is found to be applicable in the present case of $LaMnO_3$ and also valid in similar doped $La_2CuO_4$ family of compounds.[13] The fitting of the impedance spectra is shown in Fig. 2b and 2c. It yields $β$ to be temperature-dependent and varying between 0.8-0.95 (Fig.3c). Therefore, the relaxation spectra appear to be varying between nearly Debye to non-Debye type with a spectrum of relaxation times ($τ$). Good fitting of the experimentally observed impedance spectra (Figs. 2b and c) by Davidson-Cole type generalized model shows that there is no need to invoke more complex models of relaxation for first-hand analysis. The relaxation could result from fluctuation in local polar domains and/or hopping of polarons. It is interesting to note that there is a sharp anomaly in $β$-$T$ pattern at $T_N$: $β$ jumps up to a near-Debye value (~0.9) at $T_N$. The steady decrease in $β$, otherwise, with temperature results from broadening in relaxation time spectrum due to rise in thermal fluctuation of domain-domain coupling. At $T_N$, however, possibly a sharp rise in polarization domain volume occurs in the absence of spin domains and hence one observes reemergence of Debye-like scenario. Finally, again the thermal fluctuation yields a steady decrease in $β$. Such an anomaly in $β$ at $T_N$ highlights a role of the spin order in influencing the polar domains and their relaxation dynamics.

It is possible to calculate the polar domain volume by using a more involved model of dielectric relaxation in heterogeneous systems.[16] In this model, the overall electrical displacement ($D$) is given by the summation of local polarization due to electrical domains and a linear coupling between polarization ($P$) and magnetization ($M$) which can yield $P$ as a function of $M$. This is because of the orbital order, which drives both the polarization $P$ and the superexchange interaction across Mn-O-Mn bond that governs the magnetization $M$. Eventually, dielectric relaxation equations can be written as a function of domain volume $\Phi$, $\beta$, $\tau$, $\Delta\varepsilon$ etc. The model and the results of fitting with the experimental relaxation patterns across $T_N$ will be published elsewhere.

We have estimated the equivalent circuit elements $R_N$ and $C_0$ as a function of temperature. In Fig. 3, the pattern of variation in $R_N$ and $C_0$ with temperature is shown. Also shown in Fig. 3 is the relaxation time ($\tau_0$) versus temperature. The plots of $C_0$ and $\tau_0$ versus temperature depict clear anomaly at $T_N$. *These are the central results of this paper*. The $\tau_0$ versus $T$ patterns turn out to be Arrhenius both at below and above $T_N$. The fitting of $\tau_0$ vs. $T$ pattern with Arrhenius equation $\tau_0 = \tau_\infty \exp(E/k_B T)$ yields the activation energy $E$ associated with relaxation; $E$ turns out to be ~362 and ~580 K at below and above $T_N$, respectively. The smaller E at below $T_N$ signifies faster fluctuation or faster hopping of polarons due to the presence of in-plane ferromagnetic order. The $T_N$ has been verified (data not shown) from dc magnetization vs. temperature measurement.

The origin of the dielectric anomaly at $T_N$ in LaMnO$_3$ is still not clearly understood. There could be few reasons: (i) resistive component of dielectric response together with Maxwell-Wagner effect can mimic the feature of a genuine multiferroic system,[17] (ii) structural transition or striction effect at $T_N$, and (iii) magnetic order driven electrical polarization as observed in TbMnO$_3$.[18] Since we have clearly observed anomaly in the capacitive component $C_0$ at $T_N$, role of the resistive part is insignificant. Also, there is no report so far on structural transition or striction effect across $T_N$ in LaMnO$_3$. Finally, it has been fairly well settled[18] that incommensurate spiral magnetic structure in TbMnO$_3$ or DyMnO$_3$ breaks both the spatial and temporal inversion symmetry which gives rise to ferroelectric polarization at the incommensurate magnetic transition ($T_{IC}<T_N$). Since, such inhomogeneous magnetic structure does not exist in LaMnO$_3$ [Ref. 19], the possibility of magnetic structure driven polarization can be ruled out. It is also noteworthy that $\varepsilon_{r0}$ in LaMnO$_3$ is reasonably high ($\varepsilon_{r0} \sim 18$-20).[4] For a genuine multiferroic system $\varepsilon_{r0}$ is ~ 25 [Ref. 20] whereas for the non-polar systems, $\varepsilon_{r0}$ varies within 1-5. Therefore, such non-negligible electrical polarization must either be driven by mixed electron-lattice effect or by purely electronic effect. We point out here that there is evidence of development of orientation polarization driven by *electronic mechanisms* in other strongly correlated electron systems such as Ruddlesden-Popper compound Pr(Sr$_{0.9}$Ca$_{0.1}$)$_2$Mn$_2$O$_7$ [Ref. 21], multiferroic BiMnO$_3$ [Ref. 22] and LuFe$_2$O$_4$ [Ref. 2] where orbital and charge order, respectively, are claimed to be responsible. Moreover, the observation of charge/orbital order driven polarization and dielectric anomaly near order-disorder transition in doped manganite has been reported in Ref. 23. All these results indicate that it is not unlikely that orbital order in LaMnO$_3$ drives electrical polarization, at least, locally. The dielectric anomaly at $T_N$, then finds a natural explanation via spin-orbital order coupling. Of course, detailed picture of how long range orbital order gives rise to polarization and whether it is purely electronic effect or mixed electron-lattice effect is remaining unclear. It is also worthwhile to mention here that there is a minor possibility of spin-charge coupling influencing the dielectric response. Although, the orbital-spin coupling is quite strong in this undoped LaMnO$_3$ system and therefore, the dielectric anomaly near $T_N$, being reported in this paper, can be considered as a signature of spin-orbital coupling, the possibility of simultaneous influence of spin-charge coupling cannot be ruled out. Further work is needed for unraveling the picture behind the orbital order driven electrical polarization in such undoped compounds. Such work will help in searching for systems where orbital order can give rise to global ferroelectric order and hence strong multiferroicity.

In summary, we report a clear dielectric anomaly around $T_N$ in LaMnO$_3$ which cannot originate either from structural transition or lattice modulation or resistive effect. On the contrary, it finds a simple explanation via spin-orbital coupling if the orbital order drives electrical polarization in LaMnO$_3$. The mechanism – electronic versus mixed electronic and lattice – behind the orbital order driven polarization, however, is still not properly understood. Our result may stimulate enumeration of the clear picture which, in turn, can pave the way for discovery of newer systems where ferroelectricity and multiferroicity are driven by orbital structure.

We acknowledge useful discussion with A. Sen, A.K. Raychaudhuri, and P. Mahadevan.

*Corresponding author; electronic address: dipten@cgcri.res.in

[1] See, for example, W. Eerenstein, N.D. Mathur, and J.F. Scott, Nature **442**, 759 (2006).

[2] N. Ikeda, H. Ohsumi, K. Ohwada, K. Ishii, T. Inami, K. Kakurai, Y. Murakami, K. Yoshii, S. Mori, Y. Horibe, and H. Kitô, Nature **436**, 1136 (2005).

[3] D.V. Efremov, J. van den Brink, and D.I. Khomskii, Nature Mater. **3**, 853 (2004).

[4] J.L. Cohn, M. Peterca, and J.J. Neumeier, Phys. Rev. B **70**, 214433 (2004); P. Lunkenheimer, V. Bobnar, A.V. Pronin, A.I. Ritus, A.A. Volkov, and A. Loidl, Phys. Rev. B **66**, 052105 (2002); A.S. Alexandrov and A.M. Bratkovsky, J. Phys.:Condens. Matter **11**, L531 (1999).

[5] J.B. Goodenough, Phys. Rev. **100**, 564 (1955).

[6] T. Hotta, S. Yunoki, M. Mayr, and E. Dagotto, Phys. Rev. B **60**, R15009 (1999); G. Khaliullin and V. Oudovenko, Phys. Rev. B **56**, R14243 (1997).

[7] L. Paolasini, R. Caciuffo, A. Sollier, P. Ghigna, and M. Altarelli, Phys. Rev. Lett. **88**, 106403 (2002).

[8] P. Mandal, B. Bandyopadhyay, and B. Ghosh, Phys. Rev. B **64**, 180405(R) (2001).

[9] T. Chatterji, F. Fauth, B. Ouladdiaf, P. Mandal, and B. Ghosh, Phys. Rev. B **68**, 052406 (2003).

[10] D. Bhattacharya, P.S. Devi, and H.S. Maiti, Phys. Rev. B **70**, 184415 (2004).

[11] P. Mondal, D. Bhattacharya, and P. Choudhury, J. Phys.:Condens.Matter **18**, 6869 (2006).

[12] M.E. Lines and A.M. Glass, *Principles and Applications of Ferroelectrics and Related Materials* (Clarendon, Oxford, 1977); J.Y. Jo, H.S. Han, J.-G. Yoon, T.K. Song, S.-H. Kim, and T.W. Noh, arxiv.org/cond-mat/0704.1053v1 (2007).

[13]T. Park, Z. Nussinov, K.R. Hazzard, V.A. Sidorov, A.V. Balatsky, J.L. Sarrao, S.-W. Cheong, M.F. Hundley, J.-S. Lee, Q.X. Jia, and J.D. Thompson, Phys. Rev. Lett. **94**, 017002 (2005).

[14]See, for example, A.K. Jonscher, *Dielectric relaxation in solids* (Chelsea, London, 1983).

[15]F. Henn, J. Vanderschueren, J.C. Giuntini, and J.V. Zanchetta, J. Appl. Phys. **85**, 2821 (1998); Y. Feldman *et al*., Phys. Rev. E **54**, 5420 (1996).

[16]See, for example, K. Asami, Prog. Polym. Sci. **27**, 1617 (2002).

[17]G. Catalan, Appl. Phys. Lett. **88**, 102902 (2006).

[18]M. Mostovoy, Phys. Rev. Lett. **96**, 067601 (2006); See also, S.-W. Cheong and M. Mostovoy, Nature Mater. **6**, 13 (2007).

[19]E.O. Wollan and W.C. Koehler, Phys. Rev. **100**, 545 (1955).

[20]See, for example, N. Hur, S. Park, P.A. Sharma, J.S. Ahn, S. Guha, and S.-W. Cheong, Nature **429**, 392 (2004).

[21]Y. Tokunaga, T. Lottermoser, Y. Lee, R. Kumai, M. Uchida, T. Arima, and Y. Tokura, Nature Mater. **5**, 937 (2006).

[22]C.-H. Yang, J. Koo, C. Song, T.Y. Koo, K.-B. Lee, and Y.H. Jeong, Phys. Rev. B **73**, 224112 (2006).

[23]S. Mercone, A. Wahl, A. Pautrat, M. Pollet, C. Simon, Phys. Rev. B **69**, 174433 (2004).

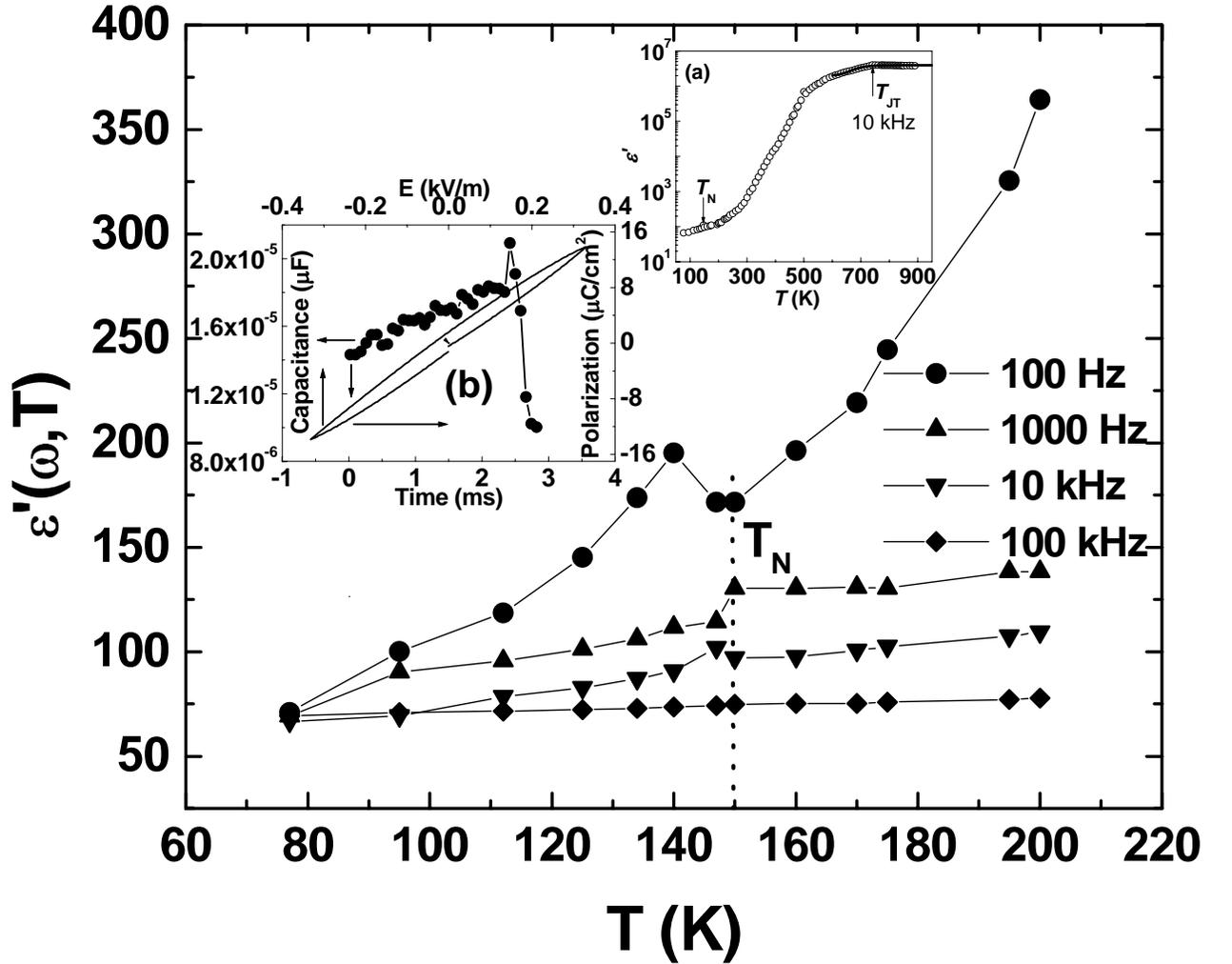

Fig. 1. Real part of the permittivity [$\varepsilon'(\omega,T)$] versus temperature (T) patterns for LaMnO$_3$ single crystals; the anomaly at T$_N$ is conspicuous. There appears to be a slight frequency dependence of the nature of the anomaly. Insets: (a) the $\varepsilon'(\omega,T)$ versus T pattern over a wide temperature range; the anomalies at T$_N$ and Jahn-Teller transition point T$_{JT}$ are conspicuous; the region near T$_{JT}$ is fitted (solid line) by a local polar domain (defect-lattice) model (b) the polarization (P) versus electric field (E) loop and the variation of capacitance with time (t) at 77 K.

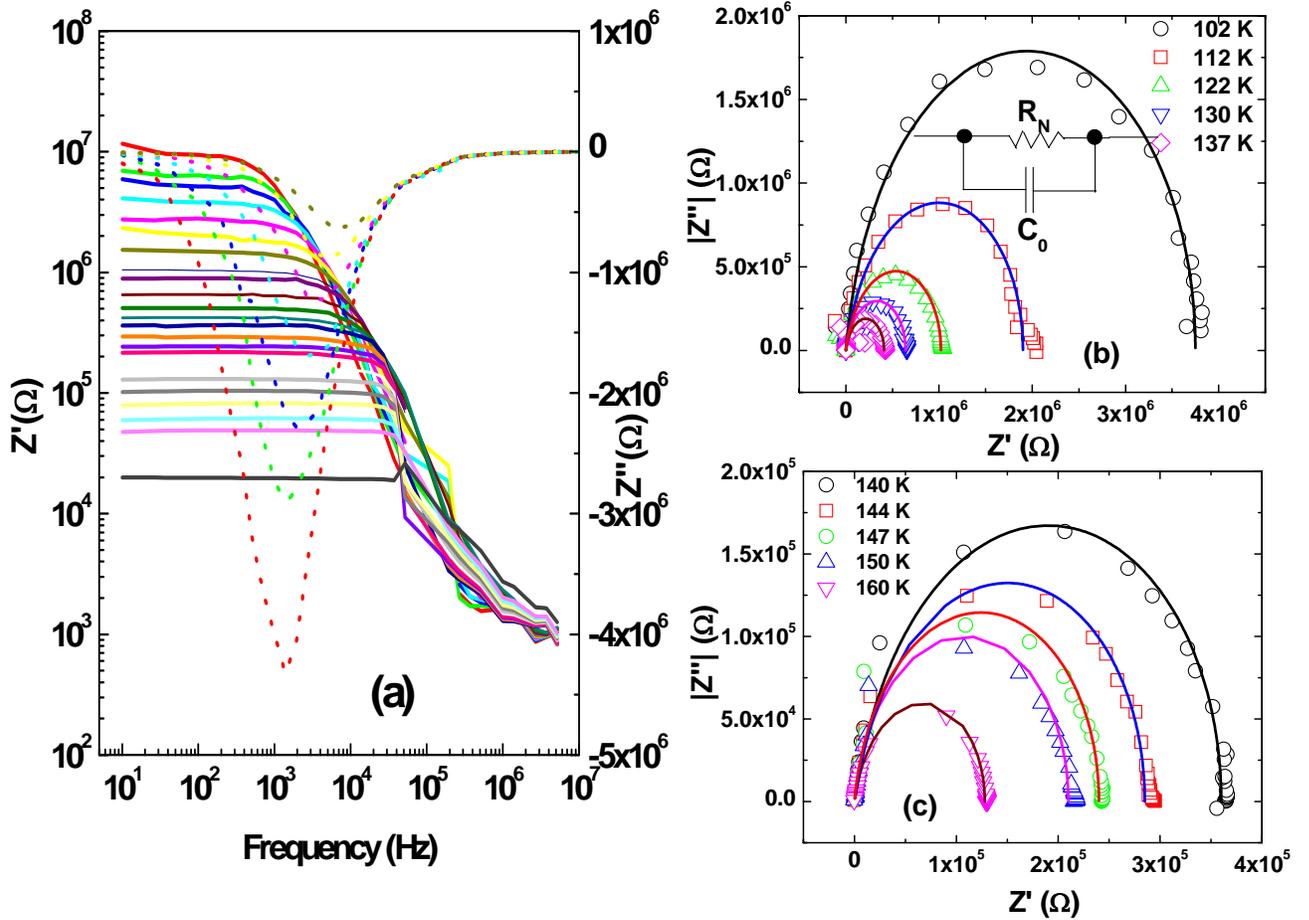

Fig. 2. (color online). (a) Real (solid line) and imaginary (dotted line) parts of the impedance [$Z(\omega)$] spectra at different temperatures; in real part the corresponding temperatures are: from top 90, 95, 98, 102, 107, 112, 115, 122, 125, 130, 134, 137, 140, 144, 147, 150, 160, 165, 170, 175, 180, 200 K and in imaginary part the corresponding temperatures are: from bottom 90, 95, 98, 102, 107, 112, 115 K; (b) and (c) Complex plane impedance patterns at below, close and above $T_N$; the solid lines have been obtained from Eq. (1) and the equivalent circuit with circuit elements is shown in the inset.

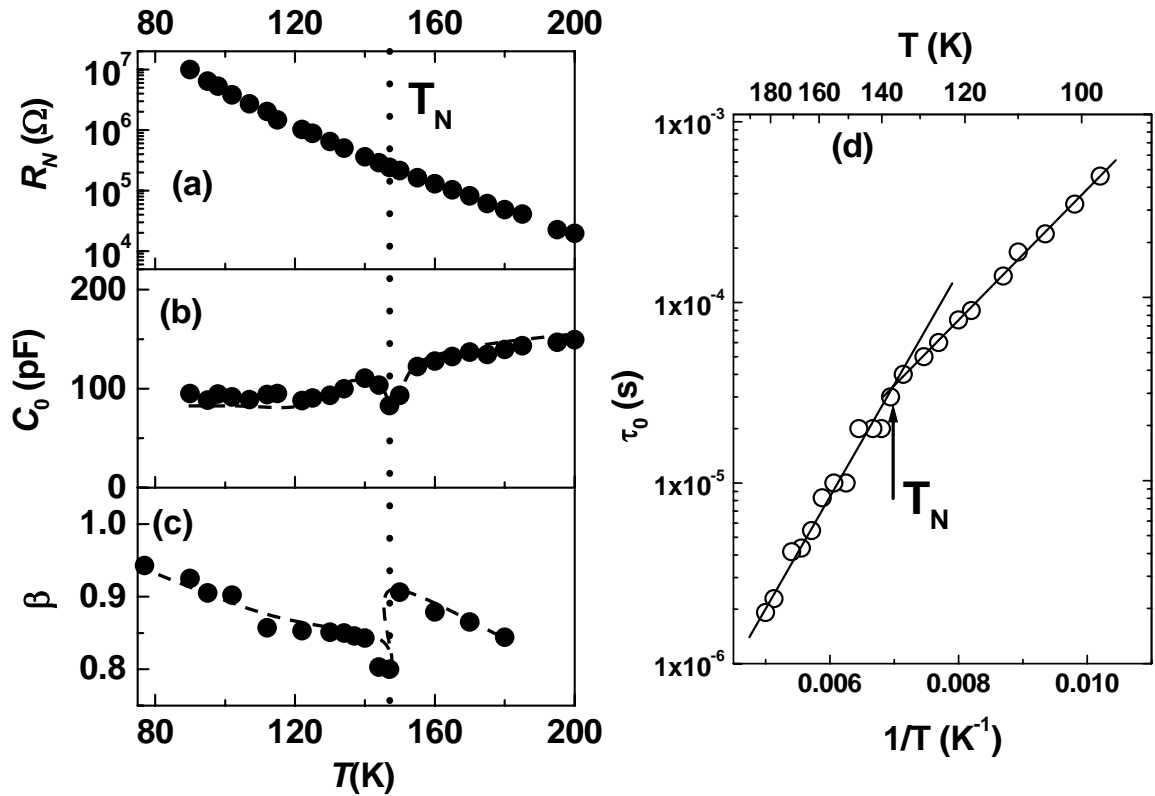

Fig. 3. The variation of equivalent circuit parameters – (a) frequency-independent resistance ($R_N$) and (b) capacitance ($C_0$) – as well as (c) the Kohlrausch exponent $\beta$ and (d) the dielectric relaxation time ($\tau_0$) with temperature across the transition point $T_N$. Dashed lines are guides to the eye.